\documentclass[12pt]{article}
\begin{document}
%%%%%%%%%%%%%%%%%%%%%%%%%%%%%%%%%%%%%%%%%%%%%%%
\newcommand{\ba}{\begin{array}}
\newcommand{\ea}{\end{array}}
\newcommand{\bea}{\begin{eqnarray}}
\newcommand{\eea}{\end{eqnarray}}
\newcommand{\edoc}{\end{document}}
\newcommand{\Pfaff}{\mbox{Pfaff}}
\newcommand{\eee}{\mbox{e}}
\newcommand{\lan}{\langle}
\newcommand{\ran}{\rangle}
\newcommand{\pal}{\partial}
\newcommand{\sta}{\stackrel}
\newcommand{\lla}{\longleftarrow}
\newcommand{\lra}{\longrightarrow}
\newcommand{\spsigma}{\ba[t]{c} \mbox{Sp} \vspace{-1ex} \cr
\mbox{$\scriptstyle{(\sigma)}$} \ea }
\newcommand{\spursigma}{\ba[t]{c} \mbox{Sp} \vspace{-1ex} \cr
\mbox{$\scriptstyle{(\sigma)}$} \ea }
%\renewcommand{\textheight}{238mm} %BOC
%\renewcommand{\textwidth}{160mm}  %BOC
%\topmargin  -23mm \oddsidemargin 1mm % laser
%%%%%%%%%%%%%%%%%%%%%%%%%%%%%%%%%%%%%%%%%%%%%%
%
%  starting the title
%
%
\phantom{A}
\vskip 5mm
\begin{center}
\section*{\Large\bf
Fermions and Correlations \\* in the Two-Dimensional Ising Model}
\vskip 3mm
{\bf V.N.~Plechko }\\
\vskip 3mm
{Bogoliubov Laboratory of Theoretical Physics,}\\
{Joint Institute for Nuclear Research, 141980 Dubna, Russia}
\end{center}
\vskip 5mm
\begin{abstract}
The anticommuting analysis with Grassmann variables is applied
to the two-dimensional Ising model in statistical mechanics. The
discussion includes the transformation of the partition function into a
Gaussian fermionic integral, the momen\-tum-space representation and the
spin-fermion correspondence at the level of the correlation functions.
\end{abstract}
\vskip 8mm

{\renewcommand{\thefootnote}{*}
\footnotetext{ \ Published in:  Physics of Particles and Nuclei,
Vol. 36, Suppl. 2, 2005, p. S203--S208. Proceedings of the Bogolyubov
Conference on Problems in Mathematical and Theoretical Physics,
September 2--6, 2004, MIRAS--JINR, Moscow--Dubna, Russia.}}
\setcounter{footnote}{0}

%
%
%   starting the text
%

There are many remarkable interrelations and analogues in studies of model
systems in quantum field theory and statistical mechanics. The
two-dimensional (2D) Ising model (2DIM) may be a good example of this kind.
In its original formulation, this is a discrete-spin lattice model of a
ferromagnet [1-7]. The important role of 2DIM in the theory of phase
transitions is due to the possibility of exact solution at all
temperatures (yet in a zero magnetic field) [1-23]. A remarkable feature is
that 2DIM can as well be reformulated in terms of free Majorana fields on a
lattice [8-15]. This can be done most effectively in terms of anticommuting
Grassmann variables and integrals [8,24]. In this report we shortly review
the application of anticommuting integrals to 2DIM (also see [18]).  The
discussion includes the transformation of the partition function $Q$ into a
Gaussian fermionic integral, the momentum-space analysis, and the
spin-fermion correspondence at the level of correlation functions.
The mirror-ordered factorization scheme we apply for translating $Q$ into
Grassmann variable integral was originated in [12-14]. For recent
developments in the theory of 2D Ising model also see [16-23].

In physical aspect, Grassmann variables are in essence the classic
(nonquantum) fermions [8,9,24]. The rules of integration over Grassmann
variables were invented by F.A. Berezin in early  60\,ies in the
quantum-field theoretical context [8]. The original motivation of his work
was in looking for a suitable fermionic counterpart for ordinary variables
of commuting analysis (bosonic fields) in path integral [8]. Since that
time, the anticommuting integral calculus has found numerous applications
in quantum field theory and condensed matter physics. Before proceeding
directly to the Ising model, we shortly remember the basic rules of
anticommuting integration.

In mathematical aspect, Grassmann variables may be viewed simply as
formal purely anticommuting fermionic symbols [8,9,24]. Given a set of $N$
independent Grassmann variables $a_1,a_2, a_3,\ldots, a_N$, one has:
$a_{i}a_{j} +a_{j}a_{i} =0\,,\,a_{j}^{\,2}=0$.  The variables may be
multiplied by each other and by complex numbers in a usual way. The linear
superpositions of Grassmann variables are again purely anticommuting, their
squares are zeros. The first important identity of the anticommuting
analysis follows by multiplying the linearly
transformed variables:
\bea
b_1\,b_2\,b_3\,....\,b_N\,=\;\det\hat{A}\,\cdot\,
a_1\,a_2\,a_3\,...\,a_N\;,\;\;\;
b_i =\sum\limits_{j=1}^{N}A_{ij}^{}a_{j}^{}\,.\;\;
\label{gra1}
\eea
Berezin's rules of the integration for one variable are [8]:
$\int da_j \cdot a_j=1,\, \int da_j \cdot 1 =0$. In multiple integral, the
differentials are again anticommuting with each other and the variables
[8]. Any natural function of Grassmann variables $a_1,a_2,\ldots,a_N$ may
be in principle represented as a finite polynomial in these variables. The
integration then reduces to the repeating application of the elementary
rules [8]. The Gaussian fermionic
integral of the first kind with integration over complex-variable (Dirac)
fermionic fields is expressed as the determinant of the matrix from the
quadratic fermionic action in the exponential:
\bea
\int\prod\limits_{j=1}^{N}\,da_{j}^{\,*}da_{j}\,\exp\Big(
\sum\limits_{i=1}^{N}\sum\limits_{j=1}^{N}a_{i}A_{ij}a_{j}^{\,*}
\Big)\,=\,\det\hat{A}\;.
\label{gra2}
\eea
The appearance of the determinant in (\ref{gra2}) may be traced back to
(\ref{gra1}). The exponential in (\ref{gra2}) is in fact a finite
polynomial due to nilpotent property of fermions. The same polynomial
follows by multiplying elementary factors $\exp\,\{a_{i}A_{ij}
a_{j}^{\,*}\} =1+a_{i}A_{ij}a_{j}^{\,*}$. The Gaussian fermionic integral
of the second kind, for real (Majorana) fermionic fields, is related to
the Pfaffian:
\bea
\int\,da_N\,...\,da_2\,da_1\,\exp\Big(
\frac{1}{2}\,\sum\limits_{i=1}^{N}\sum\limits_{j=1}^{N}
a_{i}A_{ij}a_{j}\Big)\,=\,\mbox{Pfaff}\,\hat{A}\,,\;\;
\label{gra4}
\eea
where the matrix $\hat{A}$ is assumed skew-symmetric, $A\,A^{T}
+A^{T}A =0$, in components, $A_{ij}+A_{ji} =0\,,\, A_{ii}^{}=0$. The number
of variables $N$ is assumed to be even. In mathematics, the Pfaffian is a
combinatorial polynomial associated with a triangular array of elements
$A_{ij}$. In given context, the triangular array is to be identified with
the above-diagonal array of the skew-symmetric matrix $\hat{A}$ from the
action.  The combinatorics of the Pfaffian is in fact identical to that of
the fermionic version of Wick's theorem.

The 2DIM can be solved exactly only for the homogeneous (translationally
invariant) lattice, when all couplings at the congruent bonds are equal.
The fermionization itself, however, can be performed equally well in the
general inhomogeneous case [12,18]. Therefore, we start here
with a generalized formulation of the model assuming arbitrary
inhomogeneous distribution of the bond coupling parameters over a
rectangular lattice net. The Ising spins, $\sigma_{mn}= \pm1$, are disposed
at the lattice sites, $mn$, with $m,n=1,...,L$, running in horizontal
and vertical directions, respectively, where $L$ is the length of the
lattice side. At final stages $N =L^2 \to\infty$. The hamiltonian is:
\bea
-\beta\,H(\sigma)=\sum\limits_{m=1}^{L}\sum\limits_{n=1}^{L}\; \Big[\;
b_{m+1n}^{\,(1)}\,\sigma_{mn}\sigma_{m+1n}+ b_{mn+1}^{\,(2)}\, \sigma_{mn}
\sigma_{mn+1}\,\Big]\,,\;\;
\label{ham1}
\eea
where $\beta=1/kT$ is the inverse temperature and $b_{mn}^{\,(\alpha)}
=\beta J_{mn}^{\,(\alpha)}$ are dimensionless bond coupling parameters. For
finite $L$, we assume free boundary conditions: $\sigma_{L+1n}
=\sigma_{mL+1}=0$.
The partition function is: $Z=\Sigma\,\exp\,(-\beta H)$,
where the sum is taken over spin configurations provided by $\sigma_{mn}
=\pm1$ at each site. A typical Boltzmann weight is: $\exp\,(b\, \sigma
\sigma') =\cosh b +\sigma\sigma'\,\sinh b\,$, since $(\sigma\sigma')
=\pm1$, so the problem may be readily reformulated in terms of the reduced
partition function:
\bea
Q=\spsigma\Big\{\,\prod\limits_{m=1}^{L} \prod\limits_{n=1}^{L}
(\,1+t_{m+1n}^{\,(1)} \sigma_{mn} \sigma_{m+1n})\,(1+t_{mn+1}^{\,(2)}
\sigma_{mn}\sigma_{mn+1})\,\Big\}\,,\;\;
\label{qss1}
\eea
where $t_{mn}^{\,(1,2)}=\tanh\,b_{mn}^{\,(1,2)}$ and $\,\mbox{Sp}_{
(\sigma)}$ stands for the correctly normalized spin averaging, such that
$\mbox{Sp}\,(1) =1$, $\,\mbox{Sp}\,(\sigma_{mn}) =0$ at each site.

To convert $Q$ into a Gaussian fermionic integral, the starting point is
the fermionic factorization of the local Boltzmann weights
\cite{ple85d,ple98}. We introduce a set of the totally anticommuting
Grassmann variables, $\,a_{mn},\, a_{mn}^{\,*},\, b_{mn},\,
b_{mn}^{\,*}\,$, a pair per bond, and write:
\bea
&&
1+t_{m+1n}^{\,(1)}\sigma_{mn}\sigma_{m+1n}
=\int\limits_{}^{}da_{mn}^{\,*}da_{mn}^{}\, \eee^{\,\textstyle\mathstrut
a_{mn}^{}a_{mn}^{\,*}}\,(1+a_{mn}\sigma_{mn})\,
(1+t_{m+1n}^{\,(1)}\,a_{mn}^{\,*}\sigma_{m+1n})\,,\;\;\;
\nonumber \\[-1ex]
&&
1+t_{mn+1}^{\,(2)}\sigma_{mn}\sigma_{mn+1}
=\int\limits_{}^{}db_{mn}^{\,*}db_{mn}^{}\, \eee^{\,\textstyle\mathstrut
b_{mn}^{}b_{mn}^{\,*}}\,(1+b_{mn}\sigma_{mn})\,(1+t_{mn+1}^{\,(2)}\,
b_{mn}^{\,*}\sigma_{mn+1})\,.
\phantom{xxx}\;\;
\label{fac1}
\eea
In a conventional notation, the bond Boltzmann weights are now presented
by doublets like $A_{mn}^{}A_{m+1n} ^{\,*}$ and $B_{mn}^{}B_{mn+1}^{\,*}$
standing under the sign of the diagonal Gaussian averaging arising by
factorization, where separable factors $A_{mn},B_{mn},A_{mn}^{*},
B_{mn}^{*}$ (to be called shortly Grassmann factors) are to be identified
from (\ref{fac1}). At the next stage, we keep in mind to group together,
over the whole lattice, the four factors with the same spin (the same index
$mn$) and to sum over $\sigma_{mn}=\pm1$ in each group of factors
independently thus passing to a purely fermionic expression for $Q$. The
Grassmann factors are in general neither commuting nor anticommuting with
each other, so a special arrangement of factors in their global products is
needed. The correspondent ordering procedure is called mirror-ordered
factorization \cite{ple85d,ple99}. The final result for the factorized
density matrix is:
\bea
Q\,(\sigma) =\int
\prod\limits_{m=1}^{L}\prod\limits_{n=1}^{L}
da_{mn}^{\,*}da_{mn}^{}db_{mn}^{\,*}db_{mn}^{}\;
\exp\,\Big\{\sum\limits_{m=1}^{L}\sum\limits_{n=1}^{L}\,\Big[\,
a_{mn}^{}a_{mn}^{\,*}+b_{mn}^{}b_{mn}^{\,*}\Big]\Big\}
\mbox{\phantom{XXXXXX}}
\cr
\times\,
\Big\{\!\sta{n}{\sta{\lra}{\prod\limits_{n=1}^{L}}}\!
\Big[\prod\limits_{m=1}^{L}(1+t_{mn}^{(1)}a_{m-1n}^{*}\sigma_{mn})
(1+t_{mn}^{(2)}\sta{\sta{m}{-\!\lra}}{b_{mn-1}^{*}\sigma_{mn}})
(1+a_{mn}^{}\sigma_{mn}) \cdot\!\prod\limits_{m=1}^{L}
\sta{\sta{m}{\lla\!\!-}}{(1+b_{mn}\sigma_{mn})}\Big]\Big\}.
\label{fac5}
\eea
The averaging over spins is a step-by-step
procedure realized at the junction of the two $m$-ordered products in
(\ref{fac5}), with given $n$. At each stage, the averaging
over $\sigma_{mn}=\pm1$ of the product of four neighbouring factors
$A_{mn}^{\,*} B_{mn}^{\,*} A_{mn}^{}B_{mn}^{}$ at the junction results
a simple polynomial even in Grassmann variables, which can be readily
transformed into a Gaussian exponential. This
corresponds to the local $mn$-term in the fermionic action in the second
line of (\ref{qab1}) below. In this way, one finally eliminates all spin
variables. The partition function then appears in the form a Gaussian
fermionic integral:

\bea
&&
Q\;=\; \int
\prod\limits_{m=1}^{L}\prod\limits_{n=1}^{L}
da_{mn}^{\,*}da_{mn}^{}db_{mn}^{\,*}db_{mn}^{}\;
\exp\,\Big\{\;\sum\limits_{m=1}^{L}\sum\limits_{n=1}^{L}\,\Big[\,
a_{mn}^{}a_{mn}^{\,*}+b_{mn}^{}b_{mn}^{\,*}\,+
\cr\cr
&&\,+\,t_{mn}^{\,(1)}t_{mn}^{\,(2)}\,a_{m-1n}^{\,*}b_{mn-1}^{\,*}
+\,(t_{mn}^{\,(1)}a_{m-1n}^{\,*}+t_{mn}^{\,(2)}b_{mn-1}^{\,*})\,
(a_{mn}^{}+ b_{mn}^{})\,+a_{mn}^{}b_{mn}^{}\,\Big]\Big\}\,.
\phantom{xxx}\;
\label{qab1}
\eea
The above expression is exact, and completely equivalent to (\ref{qss1}),
assuming free boundary conditions for fermions:  $a_{0n}^{\,*}
=b_{m0}^{\,*}=0$. The problem is now reformulated as a theory of free
fermions on a lattice. For further possible transformations of the
integral (\ref{qab1}) and the continuum-limit interpretation of the model
also see \cite{ple98,ple99}.

For the homogeneous (translationally-invariant) lattice, we put $t_{mn}^{\,
(1,2)} =t_{1,2}$, for definitions see (\ref{qss1}). The integral
(\ref{qab1}) can then be explicitly evaluated by passing to the momentum
space for fermions, which results Onsager's expression for the free energy
per site of infinite lattice (as $L^2\to\infty$). In the momentum space,
the integral (\ref{qab1}) becomes:

\bea
Q\;=\; \int \prod\limits_{p=0}^{L-1}\prod\limits_{q=0}^{L-1}
da_{pq}^{\,*}da_{pq}^{}db_{pq}^{\,*}db_{pq}^{}\;
\exp\,\;\Big\{\;\sum\limits_{p=0}^{L-1}\sum\limits_{q=0}^{L-1}\,
\Big[\,a_{pq}^{}a_{pq}^{\,*} + b_{pq}^{}b_{pq}^{\,*}
+ a_{pq}^{}b_{L-pL-q}^{}\,+
\nonumber \\
+\, t_{1}t_{2}\,\eee^{\,i\,\frac{2\pi p}{L}-
\,i\,\frac{2\pi q}{L}}\,a_{pq}^{\,*}b_{L-pL-q}^{\,*} +
(t_1\,\eee^{\,i\,\frac{2\pi p}{L}}\,a_{pq}^{\,*} +t_2\,\eee^{\,i\,
\frac{2\pi q}{L}}b_{pq}^{\,*})\,(a_{pq}^{} + b_{pq}^{})\,
\Big]\,\Big\}\;,\;\;\;\;\;\;\;\;
\label{qap1}
%\nonumber\\
\eea
where $a_{pq}^{}, a_{pq}^{\,*}, b_{pq}^{}, b_{pq}^{\,*}$ are the new
variables of integration introduced by discrete Fourier substitution with
periodic boundary conditions (we may change boundary
conditions into most suitable form in view of final limit $L^2\to\infty$
which abolishes boundary effects). After a proper symmetrization of the
fermionic sum with respect to $p,q \leftrightarrow L-p,L-q$, the
integral decouples into a product of elementary integral factors with
four pairs of the conjugated fermionic variables per $pq$ site. These
integral factors, can be readily evaluated by the standard rules of
fermionic integration and their product gives the squared partition
function \cite{ple88,ple99}. The result is:

\bea
Q^{\,2}\,=\,\prod\limits_{p=0}^{L-1}\prod\limits_{q=0}^{L-1}
\,\Big[\,(1+t_{1}^{\,2})(1+t_{2}^{\,2})-2t_1(1-t_{2}^{\,2})\,
\cos\frac{2\pi p}{L}- 2t_2(1-t_{1}^{\,2})\,\cos\frac{2\pi
q}{L}\,\Big]\,,\;\;
\label{qqf1}
\eea

\noindent
which is the exact solution for $Q^{\,2}$ in the limit $L^{\,2}\to\infty$.
The correspondent free energy per site, $-\beta f_{Q} =\frac{1}{L^2}\log
Q\,|\,_{L^2 \to\infty}\,$, then follows:

\bea
-\beta f_{Q} =\frac{1}{2}\int\limits_{0}^{2\pi}\int\limits_{0}^{2\pi}
\frac{dp}{2\pi} \frac{dq}{2\pi}\ln\,\Big[(1+t_{1}^{2})(1+t_{2}^{2})
-2t_{1}(1-t_{2}^{2})\cos p - 2t_{2}(1-t_{1}^{2})\cos q\,\Big]\,.\;\;
\label{ftt1}
\eea

\noindent
Respectively, the true free energy per site, for $Z=\Sigma\,\exp\,(-\beta
H)$, can be recalculated from $Z{=}(2\cosh b_{1}\cosh b_2)^{L^2}Q$,
and we gain:
\bea
-\beta f_{Z} =\ln 2  +\, \frac{1}{2}\,\int\limits_{0}^{2\pi}
\int\limits_{0}^{2\pi}\frac{dp}{2\pi}
\frac{dq}{2\pi}\ln\,\Big[\cosh 2b_1\cosh 2b_2
-\sinh 2b_1\cos p -\sinh 2b_2\cos q\;\Big]\,,\;\;\;
\label{fzz1}
\eea

\noindent
which is the famous Onsager's result, see Eq.~(108) in \cite{ons44}. From
the exact solution for the free energy, it follows that the point of phase
transition in the ferromagnetic case is given by the condition
$1{-}t_1{-}t_2 {-}t_1\,t_2=0$, or $\sinh 2b_1\cdot \sinh 2b_2=1$.  The
singularity in the specific heat near $T_c$ appears to be logarithmic. For
further details on the thermodynamics of the two-dimensional Ising
model see \cite{huang63,mac73}.

%
%  CORRELATIONS
%

In general, the studies of the
correlation functions are important in physical applications and in the
theory of the critical point. Besides, the perturbative expansions produce
the correlations.
The fermionic interpretation of the two-spin correlation function in
two-dimensional Ising model is also possible but is in fact not so
simple since it is nonlocal.\footnote{ \ The nonlocal feature in the
spin-fermion correspondence at the level of partition function in 2DIM is
somewhat masked since the Ising spins in (\ref{ham1}) and, respectively,
fermions in (\ref{qab1}), are interacting only with nearest neighbours. The
nonlocality can also be related more directly to fermionic algebra
(anticommutativity) if one starts to derive the fermionic representation
for the spin-spin correlator from the factorized representation for the
density matrix (\ref{fac5}). This version is not considered in more detail
in present discussion.} The general statement is that spin-spin correlator
$\left< \sigma_{mn} \sigma_{m'n'} \right>$ is always representable as
Toeplitz like determinant of large size which elements are all expressible
in terms of the fermionic Green's functions like $\left<a_{mn}^{}
a_{m'n'}^{\,*} \right>$. This is in a precise agreement with the results of
previous combinatorial studies of correlations in 2DIM  \cite{mpw63,mac73}
(the difference is that we use anticommuting algebra instead of ordinary
combinatorics in due course of derivation). The spontaneous magnetization
then follows via the Szego-Kac theorem \cite{mpw63,mac73}.

There are no magnetic field source terms in the fermionic action in
(\ref{qab1}), but at the spin level it is possible to express the two-spin
function like $\left<\sigma_{0} \sigma_{R}\right>$ in terms of
a string of local perturbations according to the rule:  $(\sigma_{0}
\sigma_{R}) =(\sigma_{0}\sigma_{1}) (\sigma_{1} \sigma_{2}) \ldots
(\sigma_{R-1} \sigma_{R})$, that follows from $\sigma_{j}^{2} =1$,
where we assume a local (lexicographic) enumeration of spins by
travelling from site $0$ to site $R$ along some path on a lattice. Now,
consider the case of two spins disposed on the same horizontal row of a
lattice at distance $R$ from each other:
\bea
\left<\sigma_{mn}\sigma_{m+Rn}\right>
=\left<(\sigma_{mn}\sigma_{m+1n})(\sigma_{m+1n}\sigma_{m+2n})\ldots
(\sigma_{m+R-1n}\sigma_{m+Rn})\right>\,. \;\;\;
\label{guu1}
\eea
In turn, each entry $\sigma_{mn}\sigma_{m+1n}$ may be expressed as a
modification of the bond coupling parameters in the partition function
since $(\sigma_{mn} \sigma_{m+1n}) (1+t_1\,\sigma_{mn} \sigma_{m+1n})
=t_{1}\, (1+t_{1}^{-1}\, \sigma_{mn}\sigma_{m+1n})$, where use is made
again from
$(\sigma_{mn} \sigma_{m+1n})^2 =+1$. The problem thus reduces, in
principle, to the evaluation of the fermionic integral (\ref{qab1})
with modified bonds along the path. In practical aspect, though, it might
be suitable to diminish the number of the modified fermionic terms as much
as possible. This may be realized by relating $\sigma_{mn} \sigma_{m+1n}$
directly to $a_{mn}a_{mn}^{\,*}$ if one notice that introducing prefactor
$a_{mn} a_{mn}^{\,*}$ under the integral in (\ref{qab1}) effectively
annihilates the weight $(1+t_1\sigma_{mn}\sigma_{m+1n})$ in the partition
function.  This can be most readily seen introducing $a_{mn}a_{mn}^{\,*}$
under the integral at the level of local fermionic factorization in
(\ref{fac1}). This results the correspondence:
\bea
a_{mn}^{}a_{mn}^{\,*}\;\; \longleftrightarrow\;\;
\frac{1}{1+t_1\sigma_{mn}\sigma_{m+1n}}\;\;
\longleftrightarrow\;\;
\frac{1-t_1\sigma_{mn}\sigma_{m+1n}}{1-t_{1}^{2}}\,.\;\;
\label{gii2b}
\eea
The inverse correspondence readily follows:
\bea
\sigma_{mn}\sigma_{m+1n}\;\; \longleftrightarrow\;\;
\frac{1}{t_1}-\frac{1-t_{1}^{2}}{t_1}\,a_{mn}^{}a_{mn}^{\,*}
=t_{1}^{-1}\,\exp\,\{-(1-t_{1}^{2})\,a_{mn}^{}a_{mn}^{\,*}\}\,,\;\;
\label{gii2c}
\eea
which enables one to obtain the two-point spin correlation function in
terms of a nonlocal fermionic string of perturbation under the averaging,
to which one can apply fermionic version of Wick's theorem. This eventually
results a Toeplitz determinant in terms of the fermionic correlations,
see (\ref{cag4}) below. It will be suitable yet to express the local terms
from (\ref{gii2c}) as linear fermionic forms. Introducing at each site a
pair of auxiliary fermions $c_{mn}, c_{mn}^{*}$, we write:
\bea
\frac{1}{t_1}+\frac{t_{1}^{2}-1}{t_1}\,a_{mn}a_{mn}^{\,*}
=\int dc_{mn}^{\,*}dc_{mn}^{}\eee^{\,
\displaystyle c_{mn}c_{mn}^{\,*}}
\Big\{\Big(c_{mn} +a_{mn}\Big)\Big(\frac{1}{t_1}\,c_{mn}^{\,*}
+\frac{t_{1}^{2}-1}{t_1}\,a_{mn}^{\,*}\Big)\Big\}\,.\;\;
\label{caa1}
\eea
By substituting into the string product (\ref{guu1}) the correspondence
(\ref{gii2c}), one obtains the correlator as a nonlocal fermionic string
average:
\bea
\left<\sigma_{mn}\sigma_{m+Rn}\right>
=\Big<\prod\limits_{\alpha=0}^{R-1}
\sigma_{m+\alpha n}\sigma_{m+\alpha+1n}\Big>
=\Big<\prod\limits_{\alpha=0}^{R-1}\Big[t_{1}^{-1}\Big]\Big(
1+(t_{1}^{2}-1)\,a_{m+\alpha n}a_{m+\alpha n}^{\,*}\Big)
\Big>_{(a)}, \;\;
\label{cag2}
\eea
while with the use of (\ref{caa1}), one gains representation:
\bea
&&
\left<\sigma_{mn}\sigma_{m+Rn}\right>_{(\sigma)}
=\left<(A_{mn}A_{mn}^{*})(A_{m+1n}A_{m+1n}^{*})
\ldots (A_{m+R-1n}A_{m+R-1n}^{*})\right>_{(a,c)}
\cr\cr
&&
=\det\,\Big\{\left<A_{m+\alpha\,n}A_{m+\beta\,n}^{*})\right>_{(a,c)}\,
\Big\}\,,\;\;\;\;\alpha,\beta =0,1,\ldots, R-1\,; \;\;
\cr\cr
&&
A_{mn} =c_{mn} +a_{mn}\,,\;\;\;\;
A_{mn}^{*}=\frac{1}{t_1}\,c_{mn}^{\,*}
+\frac{t_{1}{^2}-1}{t_1}\,a_{mn}^{\,*}\,,  \;\;\;\;
\cr\cr
&&
\left<A_{m+\alpha n}^{}A_{m+\beta n}^{*}\right>\neq 0\,,\;\;\;
\left<A_{m+\alpha n}^{}A_{m+\beta n}^{}\right>
=\left<A_{m+\alpha n}^{*}A_{m+\beta n}^{*}\right>=0\,.\;\;
\label{cag4}
\eea
That the correlator in the first line appears to be the determinant (but
not the Pfaffian) follows from the selection rules outlined in the last
line of (\ref{cag4}). Due to these rules, which in turn are based on
$\big<a_{mn}a_{m'n}\big> =\big<a_{mn}^{*}a_{m'n}^{*}\big> =0$ for the
correlators on the same horizontal row, the Pfaffian recursion of
Wick's theorem reduces to the determinantal recursion.\footnote{ \
The equations
$\left<a_{mn}a_{m'n}\right> =\left<a_{mn}^{*} a_{m'n}^{*} \right> =0$
follow by a direct calculation via the momentum space, cf.
(\ref{caf1})-(\ref{gau2}). The property that $\left< a_{mn} a_{m'n} \right>
=\left<a_{mn}^{*} a_{m'n}^{*} \right> =0$ is a particular feature of a
situation under the consideration, being valid only for the fermions
disposed on the same horizontal row. In a more general case, for spins at
arbitrary positions, the result will be the Pfaffian (it is known though
that the squared Pfaffian is again the determinant).}
The determinant that appears in (\ref{cag4}) is Toeplitz determinant of
$R\times R$ matrix with matrix element $A(\alpha-\beta) =\big<A_{m+\alpha
n} A_{m+\beta n}^{*} \big>$, for which in what follows we rather assume the
notation $A(m-m') =\big<A_{mn}^{} A_{m'n}^{*}\big>$. In terms of the
original fermions:
\bea
\left<A_{mn}^{}A_{m'n}^{*}\,\right>
=\frac{1}{t_1}\,\delta_{mm'} -\frac{1-t_{1}^{2}}{t_1}\,
\left<a_{mn}^{}a_{m'n}^{*}\right>\,. \;\;
\label{cag1}
\eea
The explicit solution for the only nonzero correlator we need follows
directly from (\ref{qab1})-(\ref{qap1}):
\bea
&&
\left<a_{mn}^{}a_{m'n'}^{\,*}\right>
=\frac{1}{L^2}\sum\limits_{p=0}^{L-1}\sum\limits_{q=0}^{L-1}
\left.\left<a_{pq}^{}a_{pq}^{\,*}\right>\,
\eee^{\,i\frac{2\pi p}{L}(m-m')+i\,\frac{2\pi q}{L}(n-n')}
\,\right|\,_{L\to\infty}
\nonumber \\
&&
=\int\limits_{0}^{2\pi}\!\int\limits_{0}^{2\pi}
\frac{dp\,dq}{(2\pi)^2}\,\frac{\,\eee^{ip(m-m')+iq(n-n')}\,
[\,|1-t_2\eee^{iq}|^2-t_{1}^{}\,\eee^{-ip}(1-t_{2}^{\,2})\,]}
{(1+t_{1}^{2})(1+t_{2}^{2})-2t_{1}^{}\,(1-t_{2}^{2})\,\cos p\,
-2t_{2}^{}\,(1-t_{1}^{2})\,\cos q\,}\,, \;\;
\label{caf1}
\eea
which we have to substitute into (\ref{cag1}), taking the case $n=n'$.
This results the integral for $\big<A_{mn} A_{m'n}^{\,*} \big>$, in which
the integration over $dq$ can be performed by residues. After some
straightforward though lengthy transformation in the numerator, and
changing $p\to 2\pi-p$, the final expression for the matrix element of the
Toeplitz determinant appears in the form:

\bea
\left<A_{mn}^{}A_{m'n}^{\,*}\right>
=\int\limits_{0}^{2\pi}\frac{dp}{2\pi}\;
\frac{
-(1-t_2-t_1\,\eee^{ip}-t_1t_2\,\eee^{ip})\,
(1+t_2-t_1\,\eee^{ip}+t_1t_2\,\eee^{ip})\,\eee^{-ip}}
{|\,(1-t_2-t_1\,\eee^{ip}-t_1t_2\,\eee^{ip})\,
(1+t_2-t_1\,\eee^{ip}+t_1t_2\,\eee^{ip})\,|}\,
\eee^{-ip(m-m')}.
\label{gau2}\;\;
\eea
Equivalently, assuming $A(m-m')\equiv \left<A_{mn}A_{m'n}^{*} \right>$, the
above expression may as well be represented in the form:
\bea
A(m-m')=\int\limits_{0}^{2\pi}\frac{dp}{2\pi}\,
\eee^{-ip(m-m')}\left[\frac{(1-t_1t_{2}^{*}\,\eee^{ip})\,
(1-t_{1}^{-1}t_{2}^{*}\,\eee^{-ip})}{(1-t_1t_{2}^{*}\,\eee^{-ip})
(1-t_{1}^{-1}t_{2}^{*}\,\eee^{+ip})}\right]^{\frac{1}{2}}
=\int\limits_{0}^{2\pi}\frac{dp}{2\pi}\,
\eee^{-ip(m-m')} c(p)\,,
\label{gau5}\;\;
\eea
where the Kramers-Wannier conjugate parameters are:
\bea
t_{1}^{*} =\frac{1-t_{1}}{1+t_{1}} =\eee^{-2b_{1}}\,, \;\;
t_{2}^{*} =\frac{1-t_{2}}{1+t_{2}} =\eee^{-2b_{2}}\,. \;\;
\label{kau1}
\eea
This is a self-dual transformation: $t_{1}^{**}=t_{1}^{},\; t_{2}^{**}
=t_{2}^{}$. The condition for the ferromagnetic critical point,
$1{-}t_1{-}t_2{-}t_1t_2 =0$, is equivalent to $t_{1}=t_{2}^{*}$ and
$t_{2}^{}=t_{1}^{*}$. The matrix element $A(m-m')$ depends only on the
difference of the indices, for which reason the correspondent determinant
is called Toeplitz determinant \cite{mpw63,mac73}. Notice that $A(m-m')$ is
yet not a symmetric function with respect to changing the sign of $m-m'$.
The last equation in (\ref{gau5}) is merely to define the momentum-density
function $c(p)$. This function $c(p)$ being in essence a phase factor
possesses interesting properties like $|c(p)|^2=1$ and
$c(-p) =1/c(p)$.\footnote{ \
It is not yet well  understood, beyond formal calculation, for what reason
one might expect for $c(p)$ to possess such particular properties. In any
case, the form of $c(p)$ is seemingly related to the special role of the
zero momentum modes (both ferromagnetic and antiferromagnetic) in path
integral (\ref{qap1}) with respect to the formation of the spontaneous
magnetization, $M$, which feature can in fact be recognized in the
explicit solution for $M$ itself, see (\ref{magi5}) below. For this
question also see \cite{ple88}.}

The correlator $\left< \sigma_{mn} \sigma_{m+Rn} \right>$ is thus expressed
in terms of Toeplitz determinant which matrix elements $A(m-m')$ are
fermionic Green's functions explicitly given in (\ref{gau5}). The similar
Toeplitz determinants typically appear in classic combinatorial treatments
\cite{mpw63,mac73}. The analysis (rather complicated) of the Toeplitz
determinants of this kind provides the all known information, at least, in
that part of it that can be derived from the first principles, about the
real-space asymptotic behaviour of the two-point spin-spin correlation
function in 2DIM \cite{mpw63,wmtb76,mac73}.

The solution for spontaneous magnetization also follows from the Toeplitz
determinant \cite{mpw63}.  In fact, the squared spontaneous magnetization,
$M^2$, is the limiting value of two-spin correlator
$\left<\sigma_{mn}\sigma_{m+Rn}\right>$ as $R\to\infty$, and may thus be
obtained by means of the Szego-Kac theorem \cite{mpw63} as the limiting
value of the Toeplitz determinant resulting in (\ref{cag4})-(\ref{kau1}) as
$R\to\infty$.  Though the Szego-Kac theorem by itself is rather a
complicated mathematical statement, as regards its justification, see a
discussion in \cite{mac73}, its application to the evaluation of $M^2$
is quite simple. The Szego-Kac
expresses the limiting value of the Toeplitz determinant in terms of the
real-space Fourier components of the function $\ln c(p)$, rather then
$c(p)$, where $c(p)$ is the Fourier image of $A(m-m')$ in notation of
(\ref{gau5}). This change from $c(p)$ to $\ln c(p)$ simplifies the
situation drastically. The spontaneous magnetization finally appears in
the form \cite{mpw63}:
\bea
M =\left[1-\left(\frac{1-t_{1}^{2}}{2t_{1}^{}}\right)^2\!
\left(\frac{1-t_{2}^{2}}{2t_{2}^{}}\right)^2\,\right]^{1/8}
=\left[1-\frac{1}{\sinh^2 2b_1\, \sinh^2 2b_2}\,\right]^{1/8}.\;\;
\label{magi5}
\eea
The above expression is valid in the ordered phase, in the domain
$\sinh 2b_1 \cdot \sinh 2b_2 >1$. It is interesting that despite of the
fact that the final expression for the spontaneous magnetization
(\ref{magi5}) is seemingly quite simple, it is even not the elliptic
integral, its derivation by any known method is few times more complicated
than the evaluation of the free energy and specific heat in the
correspondent approach. The fermionic analysis, though simplifies
combinatorial treatment, demonstrates the same feature. Respectively, the
problem of a better physical understanding of the ordering phenomena in
2DIM in terms of fermions still remains. What is ordered in two-dimensional
Ising model in terms of fermions? In this respect, it might be worth
mentioning that there are two other important problems with fermions
in two dimensions, the quantum Hall effect and the high-$T_c$
superconductivity in copper oxides. Finally, the two-dimensional Ising
model, within the fermionic interpretation, may be considered in a common
range with other typical problems in quantum statistics and quantum field
theory.

%%%%%
%
% \input{1-b-5bi4.tex}
%

\end{document}